\documentclass[english]{article}
\usepackage[T1]{fontenc}
\usepackage[utf8]{inputenc}
\usepackage[margin=1in]{geometry}
\usepackage{float}
\usepackage{amsmath}
\usepackage{graphicx}

\makeatletter

\providecommand{\tabularnewline}{\\}
\floatstyle{ruled}
\newfloat{algorithm}{tbp}{loa}
\providecommand{\algorithmname}{Algorithm}
\floatname{algorithm}{\protect\algorithmname}

\usepackage[cache=true,cachedir=mintedpip,frozencache=true]{minted}
\usepackage{xcolor}
\usepackage{url}
\usepackage{authblk}
\definecolor{LightGray}{gray}{0.9}
\usepackage{booktabs}

\makeatother

\usepackage{babel}

\providecommand{\keywords}[1]
{
	\small	
	\textbf{\textit{Keywords---}} #1
}

\begin{document}
\title{Point in polygon calculation using vector geometric methods with application to geospatial data.}
\author[1]{Eyram Schwinger}
\author[1]{Ralph Twum}
\author[1]{Thomas Katsekpor}
\author[2]{Gladys Schwinger}
\affil[1]{Department of Mathematics, University of Ghana}
\affil[2]{Institute of Environment and Sanitation Studies, University of Ghana}
\maketitle
\begin{abstract}
	In this work, we designed algorithms for the point in polygon problem based on the ray casting algorithm using equations from vector geometry. The algorithms were implemented using the python programming language. We tested the algorithm against the point in polygon algorithms used by the shapely (and by extension geopandas) library and the OpenCV library using points from the google Open Buildings project. Our algorithm in pure python performed much better than the shapely implementation. It also performed better than the OpenCV implementation when combined with the Numba optimization library. We also performed simulations to verify that our algorithm performance was of the order $\mathcal{O}(n).$ 
\end{abstract}


\subsection*{Introduction}

The point in polygon problem is a fundamental geometric problem with a wide range of applications in computer graphics, painting simulation, robotics, geosciences, remote sensing, and geographic information systems. Therefore, efficient numerical algorithms for solving point in polygon computations are essential. 

Many algorithms have been proposed (for instance in \cite{pipmethod1,pipmethod2,pipmethod3,pipmethod4,pipmethod5,pipmethod6}).
Some of the methods include the \emph{winding number method} \cite{pipmethod1,pipmethod3,pipmethod4,pipmethod6}, which measures the number of times a polygon encloses a point, the
\emph{sum of angles method} \cite{pipmethod5,pipmethod6}, where the sum of the interior angles of the polygon is used, and the \emph{ray casting
method} \cite{pipmethod5,pipmethod2}. The ray casting method involves
shooting a ray from the point in question in any direction and counting
how many times the ray intersects the boundaries of the polygon. 
Most of these methods assume the polygon is convex.

Several of the algorithms are computationally expensive and are sensitive to rounding and truncation errors.
\cite{pipmethod6}
states that the time complexities of the preprocessing step ranges
from $\mathcal{O}(n)$ to $\mathcal{O}(n^{2}).$ 
In this paper, we design an algorithm for point in polygon computation
using vector geometric methods, with the advantage that the algorithm
can take advantage of processor parallelization. 

The work is broken down in the following sections: The method section where we develop our algorithm from the vector geometric definition of the line. In the implementation section, we write the python code for the algorithm developed in methods then in the simulation section we test the algorithm on real world data and against other algorithms.

\subsection*{Method}

Our proposed algorithm is based on the ray casting method. Suppose we want
to determine whether a point $P(x,y)$ lies within a closed polygon
$R$ with vertices $R_{0},R_{1},\ldots,R_{n}$, where $R_{0}=R_{n}.$
In the ray casting method, a ray $l$ is drawn moving out from the
point $P$ in a specific direction. The point $P$ lies inside the polygon
$R$ if $l$ intersects the edges of $R$ an odd number of times.
In vector geometry, one way of writing out the equation of a line
is the normal form. Let $\boldsymbol{n}$ be a normal vector to the
line. Note that a line in the 2D plane has two normals going in opposite
directions that can be denoted $\boldsymbol{n}_{1}$ and $\boldsymbol{n}_{2}$
with $\boldsymbol{n}_{1}=-\boldsymbol{n}_{2},$ so $\boldsymbol{n}$
can be either $\boldsymbol{n}_{1}$ or $\boldsymbol{n}_{2}.$ Assuming
$\boldsymbol{a}$ is a known point on the line and $\boldsymbol{r}$
is any point on the line, the equation of the line can be written
as 
\begin{equation}
\left(\boldsymbol{r}-\boldsymbol{a}\right)\cdot\boldsymbol{n}=0.\label{eq:normal-line}
\end{equation}
 
\begin{figure}[bh]
\begin{centering}
\includegraphics[scale=0.5]{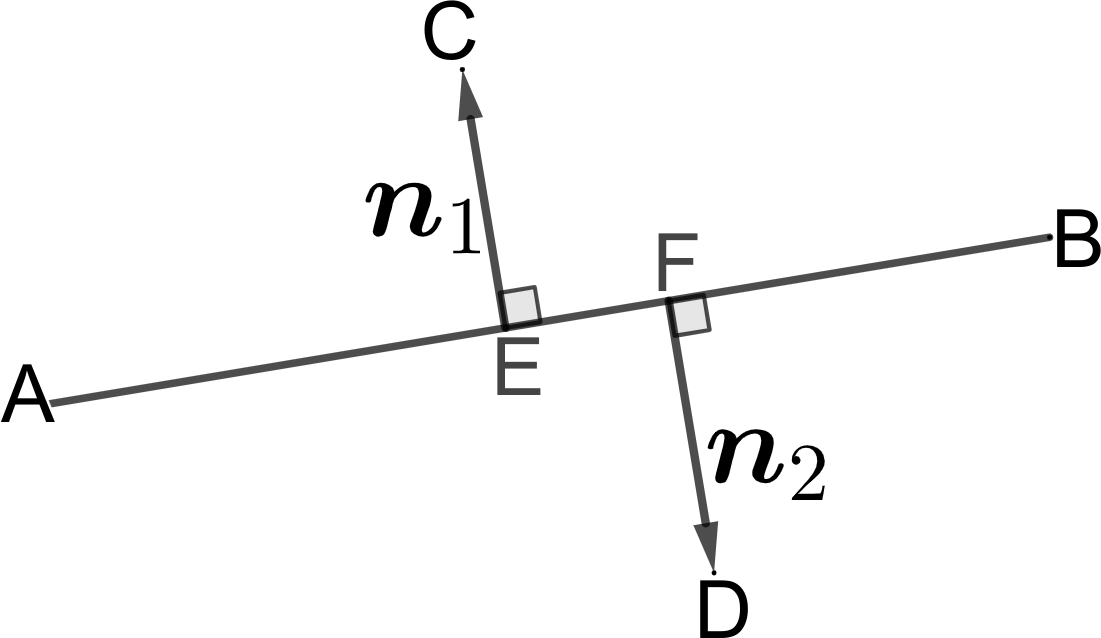}
\par\end{centering}
\caption{A line in the plane showing the two normals $\boldsymbol{n}_{1}$
and $\boldsymbol{n}_{2}$ \label{fig:Line}}

\end{figure}
Every point on the line satisfies equation \ref{eq:normal-line}.
Points that do not lie on the line do not satisfy the equation, hence
$\left(\boldsymbol{r}-\boldsymbol{a}\right)\cdot\boldsymbol{n}\ne0$
with points on opposite sides of the line having different signs.
This is illustrated in Figure \ref{fig:Line}. The point $C$ is on
one side of the line while the point $D$ is on the opposite side.
Using the concept of perpendicular distance, the position vectors
of the points $C$ and $D$ can be written as $\boldsymbol{c}=\boldsymbol{e}+\lambda\boldsymbol{\hat{n}}_{1}$
and $\boldsymbol{d}=\boldsymbol{f}+\mu\boldsymbol{\hat{n}}_{2},$
where $\hat{\boldsymbol{n}}$ represents the unit vector in the direction
of $\boldsymbol{n},$ and $\lambda$ and $\mu$ are positive constants.
Let $\boldsymbol{n}=\boldsymbol{n}_{1}.$ Then $\boldsymbol{c}=\boldsymbol{e}+\lambda\boldsymbol{\hat{n}}$
and $\boldsymbol{d}=\boldsymbol{f}-\mu\boldsymbol{\hat{n}}.$ Substituting
$\boldsymbol{c}$ and $\boldsymbol{d}$ into the equation of the line
gives 
\begin{align*}
\left(\boldsymbol{c}-\boldsymbol{a}\right)\cdot\boldsymbol{n} & =\left(\boldsymbol{e}+\lambda\boldsymbol{\hat{n}}-\boldsymbol{a}\right)\cdot\boldsymbol{n}\\
 & =\left(\boldsymbol{e}-\boldsymbol{a}\right)\cdot\boldsymbol{n}+\lambda\boldsymbol{\hat{n}}\cdot\boldsymbol{n}\\
 & =\lambda\boldsymbol{\hat{n}}\cdot\boldsymbol{n}.
\end{align*}
Now $\left(\boldsymbol{e}-\boldsymbol{a}\right)\cdot\boldsymbol{n}=0$,
since the points $A$ and $E$ lie on the line. Similarly
\begin{align*}
\left(d-\boldsymbol{a}\right)\cdot\boldsymbol{n} & =\left(\boldsymbol{f}-\mu\boldsymbol{\hat{n}}-\boldsymbol{a}\right)\cdot\boldsymbol{n}\\
 & =\left(\boldsymbol{f}-\boldsymbol{a}\right)\cdot\boldsymbol{n}-\mu\boldsymbol{\hat{n}}\cdot\boldsymbol{n}\\
 & =-\mu\boldsymbol{\hat{n}}\cdot\boldsymbol{n}.
\end{align*}
In our work, we choose to shoot a ray from the point $P$ with direction
vector $\boldsymbol{d}=(1,0).$ This is a ray shot horizontally and
to the right of the point $P.$ Then the normal vectors to the line
are $\boldsymbol{n}=(0,\pm1).$ We choose the vector $\boldsymbol{n}=(0,1).$
Since the source of the ray is $P,$ the vector $\boldsymbol{p}$
is a point on the line. This gives the equation of the line $l:\left(\boldsymbol{r}-\boldsymbol{p}\right)\cdot\boldsymbol{n}=0.$
Thus the line is the locus of points $\boldsymbol{r}\left(x,y\right)$
such that $\left(\boldsymbol{r}-\boldsymbol{p}\right)\cdot\boldsymbol{n}=0.$
This equation reduces to
\begin{align*}
\left(\boldsymbol{r}-\boldsymbol{p}\right)\cdot\boldsymbol{n} & =0\\
(x-p_{x},y-p_{y})\cdot\left(n_{x},n_{y}\right) & =0\\
\left(x-p_{x}\right)n_{x}+\left(y-p_{y}\right)n_{y} & =0\\
\left(x-p_{x}\right)\left(0\right)+\left(y-p_{y}\right)\left(1\right) & =0\\
y-p_{y} & =0.
\end{align*}
For the first constraint which is looking for ray crossings, if the
ray $l$ crosses the polygon $R$ at the edge connecting the vertices
$R_{i}$ and $R_{i+1},$ then the points $R_{i}$ and $R_{i+1}$ will
be on opposite sides of the ray $l,$ hence will have opposite signs.
Let $f\left(\boldsymbol{r}\right)=\left(\boldsymbol{r}-\boldsymbol{p}\right)\cdot\boldsymbol{n}=r_{y}-p_{y}.$
Hence $f\left(\boldsymbol{r}_{i}\right)\cdot f\left(\boldsymbol{r}_{i+1}\right)\le0$
if the ray crosses the edge $R_{i},R_{i+1}$, otherwise $f\left(\boldsymbol{r}_{i}\right)\cdot f\left(\boldsymbol{r}_{i+1}\right)>0.$
This test will however test for all edge crossings of the ray to the
left and right of $P$. 

The second constraint of the ray being cast towards the right from
the point $P$ will be enforced by testing for the vertices with $x-$components
greater or equal to the $x-$component of the point $P;$ $r_{x_{i}}\ge p_{x}.$
We tested this contraint in two ways:
\begin{enumerate}
\item Assume the points $R_{i}$ and $R_{i+1}$ are the endpoints of the
edge where the ray cast from $P$ intersects the polygon $R.$ The
constraint is satisfied if $P$ is on the left of the line $R_{i}R_{i+1}.$
To test which side of the line the point $P$ lies, define the direction
vector of the line $\boldsymbol{d}=\boldsymbol{r}_{i+1}-\boldsymbol{r}_{i}=(d_{x},d_{y}),$
then $\boldsymbol{n}=\left(-d_{y},d_{x}\right)$ or $\boldsymbol{n}=\left(d_{y},-d_{x}\right).$
Since we want to test which side of the line $P$ lies on, we choose
$\boldsymbol{n}$ to be the vector that forms an acute angle with
the vector $\left(1,0\right).$ We obtain the equation of the line
to be $l_{2}:\left(\boldsymbol{r}-\boldsymbol{r}_{i}\right)\cdot\boldsymbol{n}=0.$
Note that the vector $\boldsymbol{r}_{i}$ can be replaced by the
vector $\boldsymbol{r}_{i+1}$ in the equation of the line $l_{2}.$
Then the point lies to the left of the edge if $l_{2}:\left(\boldsymbol{p}-\boldsymbol{r}_{i}\right)\cdot\boldsymbol{n}\le0.$
\item The vector form of the equation of the line is given as $\boldsymbol{r}=\boldsymbol{r}_{i}+\lambda\boldsymbol{d}.$
This means that
\begin{align}
(x,y) & =(x_{i},y_{i})+\lambda(d_{x},d_{y})\label{eq:lamda_eq1}\\
x & =x_{i}+\lambda d_{x}\\
y & =y_{i}+\lambda d_{y}.
\end{align}
Since the ray is a horizontal line passing through the point $P$
we know that the $y-$component of the point of intersection between
the edge and the ray is $p_{y}.$ This gives 
\begin{equation}
\lambda=\dfrac{p_{y}-y_{i}}{d_{y}}.
\end{equation}
 We then substitute $\lambda$ to get $x=x_{i}+\lambda d_{x}.$ The
ray is then understood to cross the edge if $x\ge p_{x}.$
\end{enumerate}
This helps us to develop the following algorithm:
\begin{enumerate}
\item Input: $p_{x},$ $p_{y},$ vertices of the polygon $R$ as a 2-dimensional
array.
\item Find the values $f_{i}=r_{y}^{i}-p_{y}$ where $r_{y}^{i}$ is the
$y-$component of $R_{i}.$
\item Calculate sgn\_chg$_{i}=f_{i}\cdot f_{i+1}.$
\item Find $i$ where sgn\_chg$_{i}\le0.$ This means there is a ray crossing
between the vertices $R_{i}$ and $R_{i+1}.$
\item For each sgn\_chg$_{i}$ found in 4, calculate
\begin{align}
\lambda & =\dfrac{p_{y}-r_{y}^{i}}{d_{y}}\label{eq:lambda}\\
x & =r_{x}^{i}+\lambda d_{x}.\label{eq:find_x}
\end{align}
\item Find $c=x\ge p_{x}.$
\item If $c$ is odd, then $R$ contains $P.$
\end{enumerate}

\subsection*{Implementation}

The defined algorithm was implemented in python. The basic implementation
is shown in Algorithms \ref{alg:Pip1} and \ref{alg:Pip2}. In both
implementations, lines 2-4 perform the border crossing test using
the sign change algorithm. Once that is done, lines 6-7 in algorithm
\ref{alg:Pip1} calculate the normal vector $\boldsymbol{n}$ to the
point pair where the sign change occured, then lines 9-12 determine
if the calculated normal is in the same direction as the vector $(1,0).$
If it is not, the vector $-\boldsymbol{n}$ replaces $\boldsymbol{n}.$
Line 14 then calculates the sign of the point $(x,y)$ in relation
to each of the lines. Since we are only shooting the rays in one direction,
line 16 counts the number of polygon boundaries that are to the right
of the point $(x,y)$ or contain $(x,y).$ In algorithm \ref{alg:Pip2},
line 6 calculates the direction vectors of the point pair $(x_{i},y_{i}),(x_{i+1},y_{i+1})$
where the sign change occured, then line 7 calculates the parameter
$\lambda$ in equation \ref{eq:lamda_eq1} using equation \ref{eq:lambda}.
Line 8 then calculates the corresponding $x$ using equation \ref{eq:find_x}.
Line 9 then counts how many of the calculated $x$ are the same or
to the right of the input $x.$

\begin{algorithm}

\caption{Point-in-polygon algorithm using constraint 1 \label{alg:Pip1}}

\begin{minted}
[
frame=lines,
framesep=2mm,
baselinestretch=1.2,
bgcolor=LightGray,
fontsize=\footnotesize,
linenos
]
{python}
def pip_test_2(x,y,poly):
    rec = (poly[:,1]) - y
    rec = where((rec[1:]*rec[:-1]<=0))[0]
    res = poly[rec]
    res2 = poly[rec+1]
    n_x =  res2[:,1] - res[:,1]
    n_y = res[:,0] - res2[:,0]
    
    for i in range(len(n_x)):
        if n_x[i] < 0:
            n_x[i] = -n_x[i]
            n_y[i] = -n_y[i]
    
    line_pos = (x-res[:,0])*n_x+(y-res[:,1])*n_y
    
    res = count_nonzero(line_pos <= 0)
    return res %2 == 1
\end{minted}
\end{algorithm}

\begin{algorithm}

\caption{Point-in-polygon algorithm using constraint 2 \label{alg:Pip2}}

\begin{minted}
[
frame=lines,
framesep=2mm,
baselinestretch=1.2,
bgcolor=LightGray,
fontsize=\footnotesize,
linenos
]
{python}
def pip_test_1(x,y,poly):
    rec = (poly[:,1]) - y
    rec = where((rec[1:]*rec[:-1]<=0))[0]
    res = poly[rec]

    d = poly[rec+1] - res
    l = (y - res[:,1])/d[:,1]
    x_new = res[:,0] + l*d[:,0]
    res = count_nonzero(x_new > x)
    return res %2 == 1
\end{minted}
\end{algorithm}

\subsection*{Simulation}

\begin{figure}
\centering{}
\includegraphics[width=0.60\textwidth]{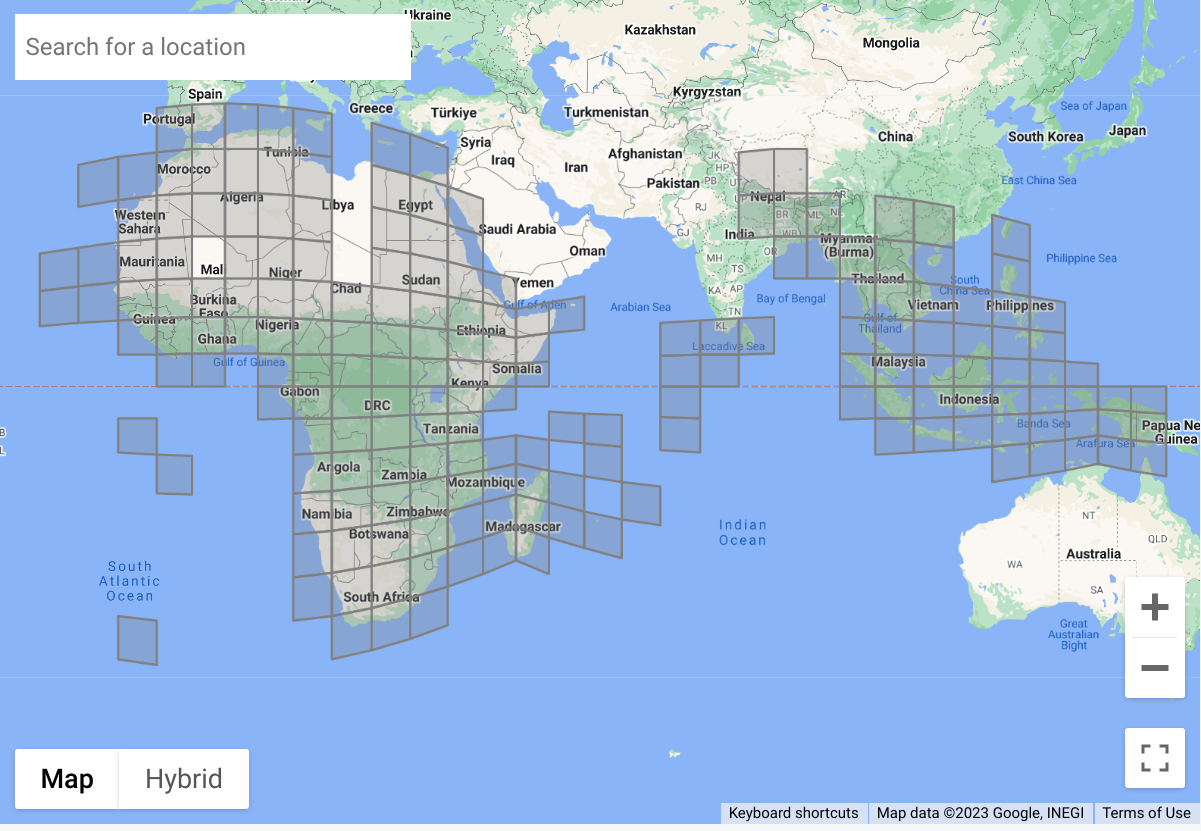}
\includegraphics[width=0.3\textwidth]{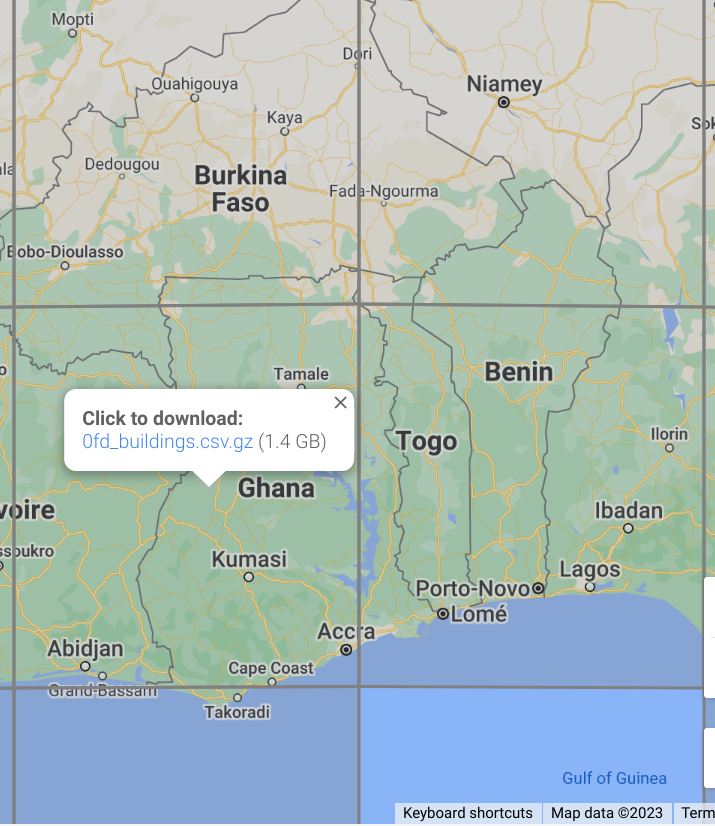}
\caption{The extent of the downloadable tiles as seen in the Google Open Buildings project. The left side shows the full coverage of the tiles while the right image shows the five tiles which were used for our work.\label{fig:google_buildings}}
\end{figure}

\begin{figure}
\caption{The map of Ghana with the points from the Open Buildings Project\label{fig:buildings}}

\centering{}\includegraphics[scale=0.5]{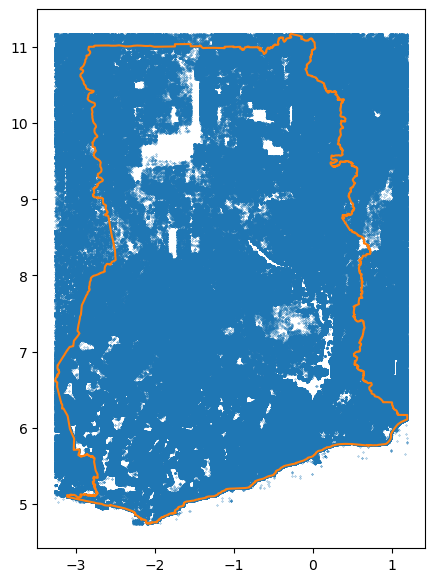}
\end{figure}

\begin{figure}
\caption{The map of Ghana showing points selected by our algorithm as being inside the polygon. The left image shows the full map with the selected points. The right image shows part of the lower boundary of the map indicated by the red rectangle at the bottom of the left image.  It shows that all the points selected as being inside the map are indeed contained in the map. \label{fig:buildings-1}}

\centering{}
\includegraphics[width=0.45\textwidth]{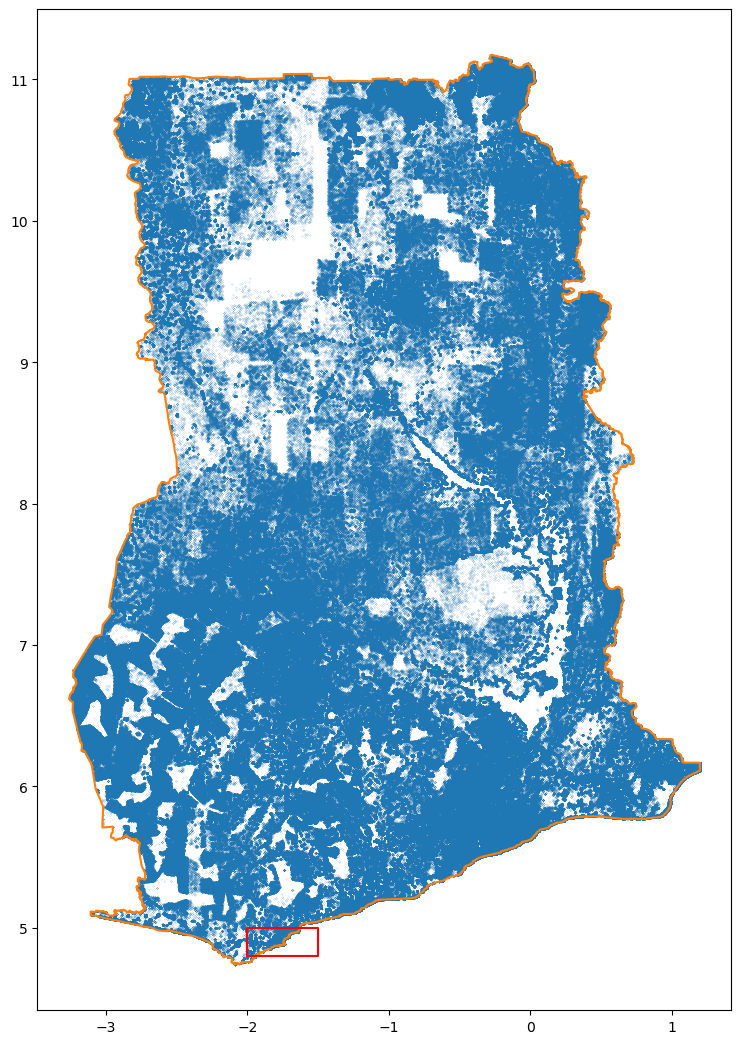}
\includegraphics[width=0.45\textwidth]{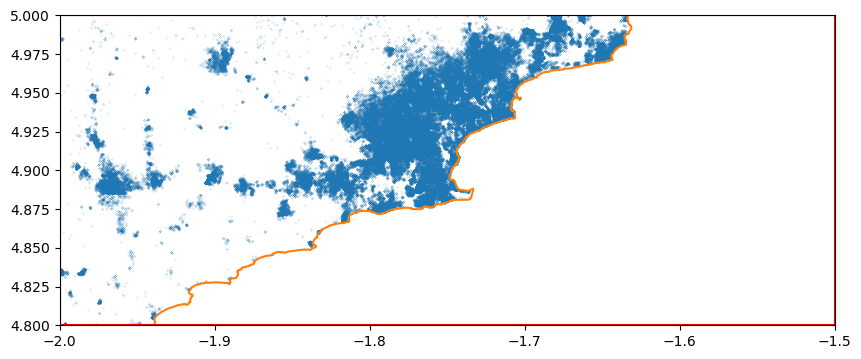}
\end{figure}

\begin{figure}
\caption{The map of Ghana showing points selected by our algorithm as being outside the polygon.\label{fig:buildings-2}}

\centering{}\includegraphics[scale=0.5]{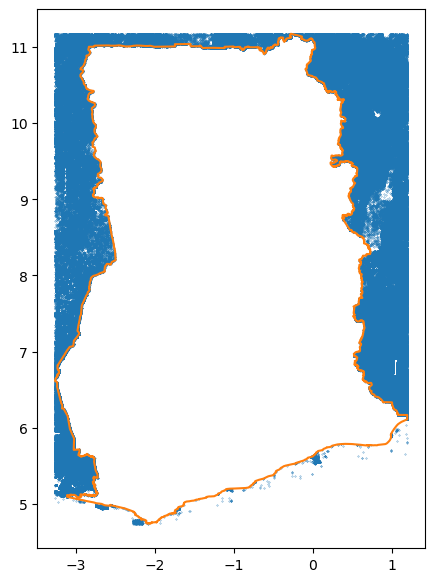}
\end{figure}

To test the algorithms, we downloaded data from the Google Open Buildings
project \cite{Sirko2021} and shapefile for the map of the Republic of Ghana from the Database of Global Administrative Areas \cite{gadm}. From \cite{Sirko2021}, we wanted data only for Ghana, so we downloaded the following cells: \texttt{0fd\_buildings},
\texttt{0e3\_buildings}, 
\texttt{11d\_buildings}, \texttt{103\_buildings}, and \texttt{0ff\_buildings} as can be seen in Figure \ref{fig:google_buildings}. The initial data contained
approximately 54 million points. Initial preprocessing to reduce the
number of points was done by restricting to only points within the
bounding box of Ghana downloaded from \cite{gadm}. 
This reduced the number of points to 17,097,823. We run our algorithm with these points as the $x,y$ points and the map of Ghana as the polygon. Figure \ref{fig:buildings}
shows the polygon with the map of Ghana and the points to be tested.
Figures \ref{fig:buildings-1} and \ref{fig:buildings-2} show the
results of all points for which our algorithm returned True and False
respectively. It can be seen that the algorithm works very well.

The algorithm was also tested against other algorithms. Our initial
simulation was done on a Dell XPS 15 9570 with an Intel i7-8750H 6
core/12 thread processor and 32GB of RAM running a Jupyter notebook with
python 3.10.8. On this setup, we tested our algorithm against the
point in polygon algorithms used by shapely \cite{shapely2007}, geopandas
\cite{kelsey_jordahl_2020_3946761} and OpenCV \cite{opencv_library}.
Table \ref{tab:execution_1} shows the execution time of our algorithm
compared to the three other algorithms. It can be seen that shapely
and geopandas took in excess of 16 hours to test their point in polygon
algorithms on the 17,097,823 points. In both shapely and geopandas
the \texttt{contains} method of the polygon and the \texttt{within}
method of the point were all tested with consistent results. Note
that geopandas uses shapely's point in polygon algorithm, hence the
similarity in execution times. It is only the OpenCV point in polygon
algorithm that was faster than our algorithm. The OpenCV library is
written in C++ and takes advantage of code optimization features.
\begin{table}
\caption{Execution time for different implementations of the point in polygon
algorithm \label{tab:execution_1}}

\begin{centering}
\begin{tabular}{|l|r|}
\hline 
Point in polygon algorithm & Execution time\tabularnewline
\hline 
\hline 
Algorithm \ref{alg:Pip1} & 18min 34s\tabularnewline
\hline 
Algorithm \ref{alg:Pip2} & 18min 30s\tabularnewline
\hline 
shapely & 16h 2min 36s\tabularnewline
\hline 
geopandas & 16h 38min 23s\tabularnewline
\hline 
OpenCV & 13min 20s\tabularnewline
\hline 
\end{tabular}
\par\end{centering}
\end{table}
We therefore combined our algorithm with numba, a high performance
python compiler \cite{lam2015numba} with both parallel and non-parallel
processing. These can be seen in Algorithms \ref{alg:Pip3} for numba
without parallelization and \ref{alg:Pip4} for numba with parallelization.
Algorithm \ref{alg:Pip2} was also tested on python 3.11 which
is claimed to have faster execution of code based on its built-in optimizations.
The results are shown in Table \ref{tab:pip_optimized_table}. From
the table it can be seen that our algorithm lends itself well to parallelization,
as the execution time was cut sharply from 18min 30s down to 3min
30s when all 12 processor threads are utilized. Even without parallelization,
there was an improvement in execution time from 18min 30s to 13min
17s which is slightly better than the execution time for the OpenCV
implementation. Again, our algorithm made some speed gains when run
in python 3.11. Our algorithm was also run on a Dell Optiplex 9020
with 8GB of RAM and an Intel i5-4590 CPU with 4 cores and 4 threads running
python 3.10.8. The results are shown in Table \ref{tab:Pip_low_spec}.
From the table it can be seen that despite the lower specifications
in processor and memory, the algorithm still performed appreciably
well. Optimizing the algorithm with numba on 4 cores produces run
times of 8min 59s and in 16min 3s without parallelization.
\begin{algorithm}
\caption{Algorithm \ref{alg:Pip2} with numba without parallelization \label{alg:Pip3}}

\begin{minted}
[
frame=lines,
framesep=2mm,
baselinestretch=1.2,
bgcolor=LightGray,
fontsize=\footnotesize,
linenos
]
{python}
@njit()
def numba_pip(pts,poly):
    pip = zeros(pts.shape[0])
    for i in prange(pts.shape[0]):
        x,y = pts[i,0],pts[i,1]
        rec = (poly[:,1]) - y
        rec = where((rec[1:]*rec[:-1]<=0))[0]
        res = poly[rec]
        d = poly[rec+1] - res
        l = (y - res[:,1])/d[:,1]
        x_new = res[:,0] + l*d[:,0]
        res = count_nonzero(x_new > x)
        pip[i] =  (res %2 == 1)
    return pip
\end{minted}
\end{algorithm}

\begin{algorithm}
\caption{Algorithm \ref{alg:Pip2} with numba with parallelization \label{alg:Pip4}}

\begin{minted}
[
frame=lines,
framesep=2mm,
baselinestretch=1.2,
bgcolor=LightGray,
fontsize=\footnotesize,
linenos
]
{python}
@njit(parallel=True)
def numbapip_p(pts,poly):
    pip = zeros(pts.shape[0])
    for i in prange(pts.shape[0]):
        x,y = pts[i,0],pts[i,1]
        rec = (poly[:,1]) - y
        rec = where((rec[1:]*rec[:-1]<=0))[0]
        res = poly[rec]
        d = poly[rec+1] - res
        l = (y - res[:,1])/d[:,1]
        x_new = res[:,0] + l*d[:,0]
        res = count_nonzero(x_new > x)
        pip[i] =  (res %2 == 1)
    return pip
\end{minted}
\end{algorithm}

\begin{figure}
\caption{Execution times in seconds for the implementation of point in polygon for different
number of points . \label{fig:Execution-times-1}}

\centering{}\includegraphics[scale=0.35]{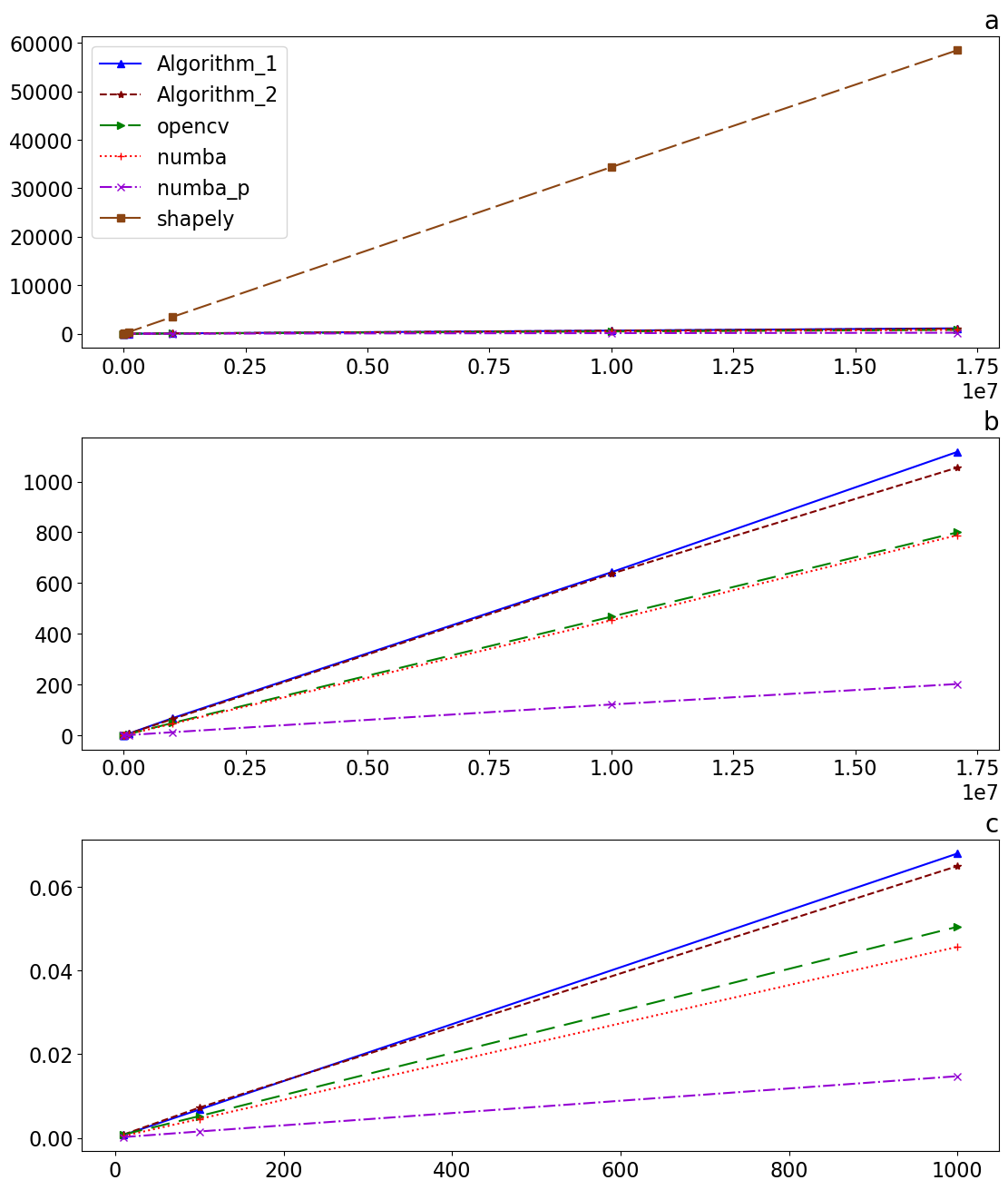}
\end{figure}

\begin{figure}

\caption{Comparison of algorithm running times using different
number of points on python 3.11 on a Dell XPS 15 9570 with an Intel
i7-8750H 6 core/12 thread processor and 32GB of RAM \label{fig:Figure-execution-times-3_11}}

\centering{}\includegraphics[scale=0.35]{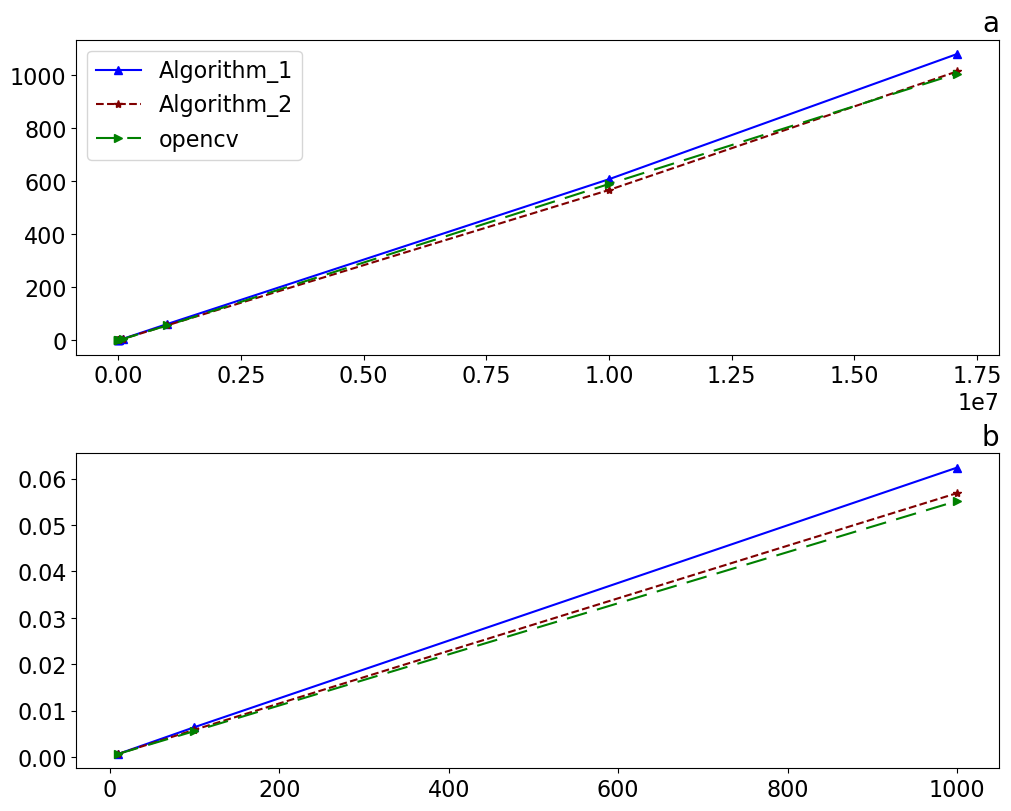}
\end{figure}

\begin{figure}

\caption{Result of overlaying Table \ref{tab:Table-projection-small} with Table \ref{tab:Table-lsq}. The lines represent the least squares fit
of the data while the markers represent the actual execution times.
\label{fig:Figure-projection-small}}

\centering{}\includegraphics[scale=0.35]{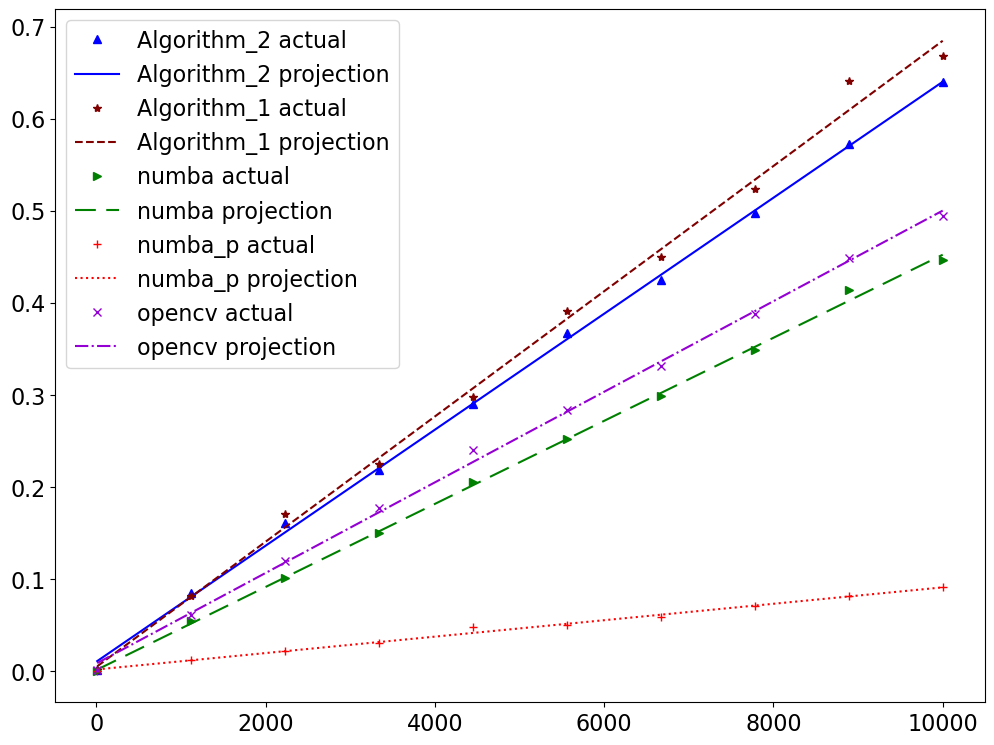}
\end{figure}

\begin{figure}
\caption{Result of overlaying Table \ref{tab:Table-projection-large} with
Table \ref{tab:Table-lsq}. The lines represent the least squares fit
of the data while the markers represent the actual execution times.
Apart from the numba parallelized implementation, the linear projection
generally bounds the actual execution times and closely follows the
lines. \label{fig:Figure-projection-large}}

\centering{}\includegraphics[scale=0.35]{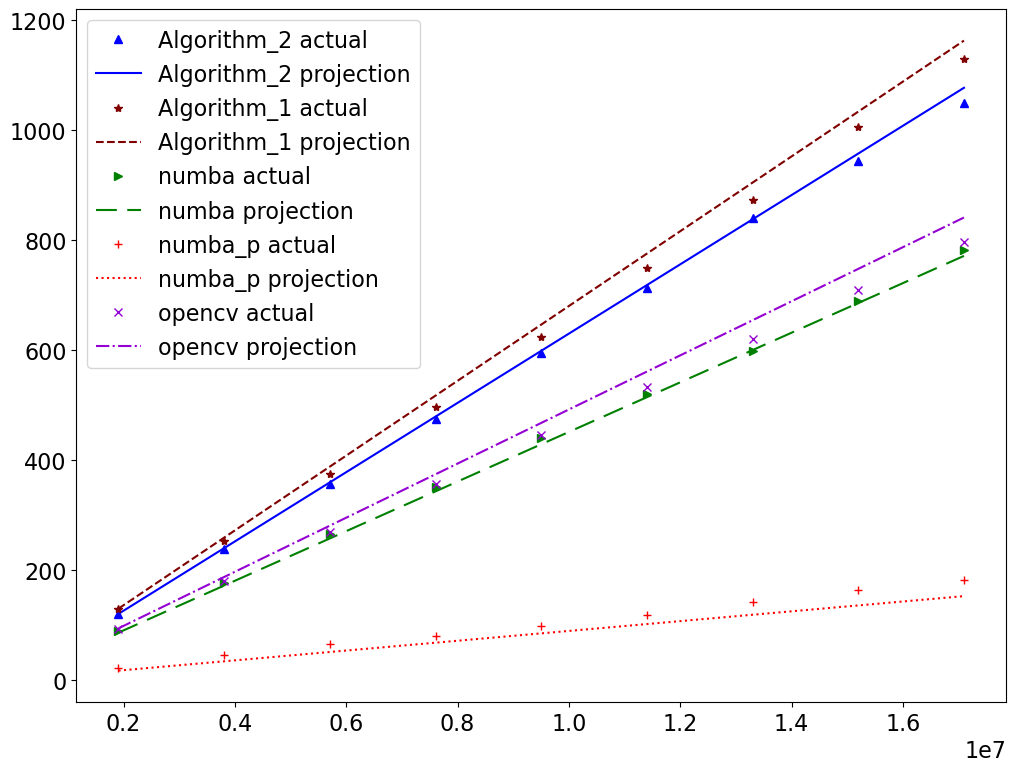}
\end{figure}

\begin{table}
\caption{Comparison of execution times of Algorithm \ref{alg:Pip2} on optimized
platforms \label{tab:pip_optimized_table}}

\centering{}%
\begin{tabular}{|l|l|r|}
\hline 
Algorithm & Platform & Execution Time\tabularnewline
\hline 
\hline 
Algorithm \ref{alg:Pip3} & numba without parallelization (python 3.10.8) & 13min 17s\tabularnewline
\hline 
Algorithm \ref{alg:Pip4} & numba with parallelization (python 3.10.8) & 3min 30s\tabularnewline
\hline 
Algorithm \ref{alg:Pip2} & python 3.11 & 16min 25s\tabularnewline
\hline 
OpenCV & python 3.10.8 & 13min 20s\tabularnewline
\hline 
OpenCV & python 3.11 & 16min 45s\tabularnewline
\hline 
\end{tabular}
\end{table}
Tables \ref{tab:Execution-times-points1}-\ref{tab:Table-projection-large}
and Figures \ref{fig:Execution-times-1}-\ref{fig:Figure-execution-times-3_11}
show the execution times of the various algorithms for different number
of points from 10 to 17,097,823. It can be seen that generally, shapely
and by extension geopandas are the worst performing methods. In Figure \ref{fig:Execution-times-1}
(a), the steep slope of the shapely algorithm makes it difficult to
see the performance of the other algorithms. The shapely algorithm
is therefore removed in the (b) and (c) parts of the figure to properly
show the performance of the other algorithms. Figure \ref{fig:Execution-times-1}
(b) shows the performance of the algorithm for logarithmically spaced
points from 10--10,000,000 and for 17,097,823 points, while the (c) part shows the
performance of the algorithm for the logarithmically spaced points
10, 100 and 1000 points which are not visible in (b). The shapely
execution time is removed from subsequent plots to allow the execution
times from the other algorithms to be visualized. It can be seen that,
consistently the numba implementation with parallelization performs
the fastest, followed by the numba implementation without parallelization,
then the OpenCV implementation, then Algorithm 2, followed by Algorithm
1. In Figure \ref{fig:Figure-execution-times-3_11}, the execution
times for the numba implementations are absent because numba was not
supported on the new python 3.11 as at the time of performing the
simulations. In Tables \ref{tab:Table-projection-small}-\ref{tab:Table-projection-large},
we show the execution times of the algorithms on a linearly spaced
scale from 10--10,000 in Table \ref{tab:Table-projection-small} and
from 1,908,647 to 17,097,823 in Table \ref{tab:Table-projection-large}.
We perform a linear least squares fit of the data in Table \ref{tab:Table-projection-small}
to find the equation of the form $ax+b=y$, where $x$ is the number
of points and $y$ is the execution time using the matrix equation
\begin{equation}
\label{mateq}
\left[\begin{array}{c|c}
\boldsymbol{x} & \boldsymbol{1}\end{array}\right]\left[\begin{array}{c}
a\\
b
\end{array}\right]=\left[\boldsymbol{y}\right].
\end{equation}
In Equation \ref{mateq}, $\boldsymbol{x}$ is the vector of the number of points that
were tested and $\boldsymbol{y}$ is the vector of execution times.
$\boldsymbol{1}$ is the vector of ones with the same length as $\boldsymbol{x}$
and $\boldsymbol{y}.$ The slopes and intercepts of the least squares
lines of best fit are shown in Table \ref{tab:Table-lsq}. Figure \ref{fig:Figure-projection-small}
shows the least squares lines and the original data used to calculate
them. Figure \ref{fig:Figure-projection-large} shows the least squares
lines overlaid on the execution times in Table \ref{tab:Table-projection-large}
to verify the linearity of the algorithms. It can be seen that the
largest deviations in execution times come from the numba algorithm
with parallelization which may be due to the calls to the multiple
processor threads. Tables \ref{tab:Absolute-error} and \ref{tab:Relative-error}
show the absolute and relative errors respectively in the execution
times projected by the least squares lines for the points in Table
\ref{tab:Table-projection-large} and the actual execution times in
the table. We can say that it is possible to use the least squares
fit of the execution times in Table \ref{tab:Table-projection-small}
with between 10 and 10,000 points to predict the execution times for
a larger number of points as contained in Table \ref{tab:Table-projection-large}.
In our case, the largest errors in absolute terms were in the opencv implementation 44.48 seconds error while the numba implementation with parallelization had the highest relative error. However and the least error was in the numba implementation without parallelization which was 11 seconds.
\begin{table}
\caption{Execution times in seconds for the implementation of point in polygon for different
number of points \label{tab:Execution-times-points1}}

\centering{}\begin{tabular}{rrrrrrr} \toprule Algorithm & Algorithm\_1 & Algorithm\_2 & opencv & numba & numba\_p & shapely \\ \midrule 10 & 0.000674 & 0.000804 & 0.000792 & 0.000455 & 0.000193 & 0.039866 \\ 100 & 0.006780 & 0.007234 & 0.005199 & 0.004525 & 0.001524 & 0.384787 \\ 1000 & 0.067971 & 0.064967 & 0.050449 & 0.045648 & 0.014742 & 3.987965 \\ 10000 & 0.674826 & 0.660946 & 0.496855 & 0.458177 & 0.131555 & 37.005583 \\ 100000 & 6.749296 & 6.394145 & 5.031192 & 4.663205 & 1.239017 & 348.450866 \\ 1000000 & 67.799494 & 64.238579 & 49.027873 & 45.881565 & 12.510255 & 3460.231409 \\ 10000000 & 642.754256 & 636.792082 & 467.686868 & 453.594473 & 121.763848 & 34361.098605 \\ 17097823 & 1117.207423 & 1055.401415 & 800.032984 & 788.796931 & 202.510374 & 58475.832875 \\ \bottomrule \end{tabular} 
\end{table}

\begin{table}

\caption{Algorithm runnung times on a Dell Optiplex 9020 computer with Intel
i5-4590 processor and 8GB of Ram \label{tab:Pip_low_spec}}

\centering{}%
\begin{tabular}{|l|r|}
\hline 
Algorithm & Execution Time\tabularnewline
\hline 
\hline 
Algorithm \ref{alg:Pip1} & 23min 57s\tabularnewline
\hline 
Algorithm \ref{alg:Pip2} & 23min 2s\tabularnewline
\hline 
Algorithm \ref{alg:Pip2} with numba without parallelization & 16min 3s\tabularnewline
\hline 
Algorithm \ref{alg:Pip2} with numba with parallelization & 8min 59s\tabularnewline
\hline 
\end{tabular}
\end{table}
\begin{table}

\caption{Comparison of algorithm running times using different
number of points on python 3.11 on a Dell XPS 15 9570 with an Intel
i7-8750H 6 core/12 thread processor and 32gb ram}

\centering{}
\begin{tabular}{rrrr} \toprule 
	Points & Algorithm\_1 & Algorithm\_2 & opencv \\ 
	\midrule 
	10 & 0.000711 & 0.000719 & 0.000720 \\ 
	100 & 0.006478 & 0.005915 & 0.005638 \\ 
	1000 & 0.062355 & 0.056861 & 0.055172 \\ 
	10000 & 0.616584 & 0.566055 & 0.548117 \\ 
	100000 & 6.154858 & 5.680473 & 5.431181 \\ 
	1000000 & 61.618874 & 56.637895 & 56.335213 \\ 
	10000000 & 606.914380 & 566.261798 & 588.910343 \\ 
	17097823 &  1078.766585 & 1013.177792 & 1005.053892 \\ \bottomrule \end{tabular}
\end{table}
\begin{table}

\caption{Comparison of algorithm running times in seconds
using different number of points on a Dell Optiplex 9020 with 8GB
of RAM and Intel i5-4590 CPU with 4 cores and 4 threads running python
3.10.8 \label{tab:Table-projection-i5}}

\centering{}\begin{tabular}{rrrrrr} \toprule Points & Algorithm\_1 & Algorithm\_2 & opencv & numba & numba\_p \\ \midrule 10 & 0.001023 & 0.003242 & 0.001362 & 0.002346 & 0.000979 \\ 100 & 0.009219 & 0.011926 & 0.006372 & 0.007208 & 0.001654 \\ 1000 & 0.086663 & 0.081737 & 0.044268 & 0.056429 & 0.015674 \\ 10000 & 0.865499 & 0.818598 & 0.441505 & 0.561596 & 0.157564 \\ 100000 & 8.678028 & 8.170648 & 4.419931 & 5.635208 & 1.825092 \\ 1000000 & 85.399976 & 82.598725 & 44.758200 & 55.967566 & 31.157190 \\ 10000000 & 850.251691 & 815.232475 & 421.485714 & 559.662381 & 309.712137 \\ 17097823 & 1459.900754 & 1387.815551 & 720.694399 & 951.991756 & 534.855275 \\ \bottomrule \end{tabular} 
\end{table}
\begin{table}

\caption{Comparison of algorithm running times in seconds
using linearly spaced points between 10 and 10000 on python 3.10.8
on a Dell XPS 15 9570 with an Intel i7-8750H 6 core/12 thread processor
and 32GB of RAM. \label{tab:Table-projection-small}}

\centering{}\begin{tabular}{rrrrrrr} \toprule Points & Algorithm\_2 & Algorithm\_1 & numba & numba\_p & opencv & shapely \\ \midrule 10 & 0.000797 & 0.000678 & 0.000455 & 0.000193 & 0.000885 & 0.040110 \\ 1120 & 0.085122 & 0.080887 & 0.053803 & 0.011901 & 0.060541 & 3.921390 \\ 2230 & 0.160281 & 0.170801 & 0.100509 & 0.021441 & 0.119104 & 7.746753 \\ 3340 & 0.218364 & 0.224381 & 0.150193 & 0.030485 & 0.176672 & 11.664765 \\ 4450 & 0.289474 & 0.298039 & 0.204831 & 0.047858 & 0.239670 & 15.478461 \\ 5560 & 0.366708 & 0.391248 & 0.251546 & 0.049396 & 0.283920 & 19.261769 \\ 6670 & 0.424220 & 0.449986 & 0.298660 & 0.058415 & 0.331706 & 23.099383 \\ 7780 & 0.497226 & 0.523772 & 0.348369 & 0.070565 & 0.387975 & 27.049710 \\ 8890 & 0.572252 & 0.640621 & 0.413792 & 0.081335 & 0.448136 & 30.752390 \\ 10000 & 0.639463 & 0.667802 & 0.446130 & 0.090945 & 0.494594 & 34.666056 \\ \bottomrule \end{tabular} 
\end{table}

\begin{table}
\caption{Comparison of algorithm running times in seconds using linearly spaced points between 1908647 and 17097823 on python 3.10.8 on a Dell XPS 15 9570 with an Intel i7-8750H 6 core/12 thread processor and 32GB of RAM.\label{tab:Table-projection-large}}

\centering{}
\begin{tabular}{llllll}
	\toprule
	Points & Algorithm\_2 & Algorithm\_1 & numba & numba\_p & opencv \\
	\midrule
	1908647 & 120.244296 & 129.822878 & 86.105501 & 17.027383 & 93.905902 \\
	3807294 & 239.848545 & 258.961195 & 171.758836 & 33.963956 & 187.311751 \\
	5705941 & 359.452793 & 388.099512 & 257.412170 & 50.900529 & 280.717599 \\
	7604588 & 479.057041 & 517.237829 & 343.065504 & 67.837102 & 374.123447 \\
	9503235 & 598.661289 & 646.376145 & 428.718838 & 84.773674 & 467.529295 \\
	11401882 & 718.265538 & 775.514462 & 514.372172 & 101.710247 & 560.935143 \\
	13300529 & 837.869786 & 904.652779 & 600.025506 & 118.646820 & 654.340991 \\
	15199176 & 957.474034 & 1033.791096 & 685.678841 & 135.583393 & 747.746839 \\
	17097823 & 1077.078283 & 1162.929413 & 771.332175 & 152.519965 & 841.152687 \\
	\bottomrule
\end{tabular}
\end{table}
\begin{table}

\caption{Result of linear least squares fit of the data in
table \ref{tab:Table-projection-small} \label{tab:Table-lsq}}

\centering{}\begin{tabular}{lrr} \toprule Algorithm & Slope & Intercept \\ \midrule Algorithm\_2 & 0.000063 & 0.010103 \\ Algorithm\_1 & 0.000068 & 0.004402 \\ numba & 0.000045 & 0.001039 \\ numba\_p & 0.000009 & 0.001607 \\ opencv & 0.000049 & 0.008094 \\ shapely & 0.003462 & 0.041119 \\ \bottomrule \end{tabular}
\end{table}
\begin{table}

\caption{Absolute error of the least squares prediction of the implementation
time using the linear coefficients in table \ref{tab:Table-lsq} and
the actual execution time in table \ref{tab:Table-projection-large}
in seconds \label{tab:Absolute-error}}

\centering{}
\begin{tabular}{lrrrrr}
	\toprule
	Points & Algorithm\_2 & Algorithm\_1 & numba & numba\_p & opencv \\
	\midrule
	1908647 & 0.260000 & 0.010000 & 2.070000 & 5.350000 & 0.980000 \\
	3807294 & 1.700000 & 5.630000 & 4.950000 & 11.700000 & 6.710000 \\
	5705941 & 2.800000 & 12.620000 & 8.540000 & 13.730000 & 11.750000 \\
	7604588 & 3.680000 & 20.350000 & 8.090000 & 11.360000 & 17.100000 \\
	9503235 & 4.500000 & 21.730000 & 11.970000 & 14.110000 & 22.460000 \\
	11401882 & 5.150000 & 26.140000 & 6.060000 & 16.990000 & 27.720000 \\
	13300529 & 2.730000 & 30.980000 & 2.080000 & 23.980000 & 33.250000 \\
	15199176 & 13.500000 & 28.640000 & 4.270000 & 27.470000 & 39.010000 \\
	17097823 & 27.740000 & 33.020000 & 11.230000 & 28.680000 & 44.480000 \\
	\bottomrule
\end{tabular}
\end{table}
\begin{table}
\caption{Relative error of the least squares prediction of the implementation
time using the linear coefficients in table \ref{tab:Table-lsq} and
the actual execution time in table \ref{tab:Table-projection-large}
in seconds \label{tab:Relative-error}}

\centering{}
\begin{tabular}{lrrrrr}
	\toprule
	Points & Algorithm\_2 & Algorithm\_1 & numba & numba\_p & opencv \\
	\midrule
	1908647 & 0.000000 & 0.000000 & 0.020000 & 0.240000 & 0.010000 \\
	3807294 & 0.010000 & 0.020000 & 0.030000 & 0.260000 & 0.040000 \\
	5705941 & 0.010000 & 0.030000 & 0.030000 & 0.210000 & 0.040000 \\
	7604588 & 0.010000 & 0.040000 & 0.020000 & 0.140000 & 0.050000 \\
	9503235 & 0.010000 & 0.030000 & 0.030000 & 0.140000 & 0.050000 \\
	11401882 & 0.010000 & 0.030000 & 0.010000 & 0.140000 & 0.050000 \\
	13300529 & 0.000000 & 0.040000 & 0.000000 & 0.170000 & 0.050000 \\
	15199176 & 0.010000 & 0.030000 & 0.010000 & 0.170000 & 0.060000 \\
	17097823 & 0.030000 & 0.030000 & 0.010000 & 0.160000 & 0.060000 \\
	\bottomrule
\end{tabular}
\end{table}

\subsection*{Discussion}

The implementation of a point in polygon computation using basic vector equations leads to a very useful algorithm which compares very well to the current algorithms that are in use. Moreover, we can scale the algorithm to higher dimensions by considering vector equations in these dimensions and modifying the algorithm accordingly. 

We used a linear least squares fit for the values in Table \ref{tab:Table-projection-small} because the values for each point appear to show a linear relationship between the number of points and the execution times of the algorithm. We can infer that the algorithm running time is of linear order, but with an extremely small slope, which allows it to scale very well. 

\bibliographystyle{plain}
\nocite{*}
\bibliography{point_in_polygon}

\end{document}